\documentclass[superscriptaddress,reprint,amsmath,amssymb,pra]{revtex4-2}

\usepackage{graphicx}% Include figure files
\usepackage{dcolumn}% Align table columns on decimal point
\usepackage{bm}% bold math
\usepackage{hyperref}% add hypertext capabilities
\usepackage{float}
\usepackage{svg}
\usepackage{physics}
\usepackage{array}
\usepackage{ulem}
\usepackage{amsmath}

\begin{document}

\title{Entanglement between a trapped ion qubit and a 780-nm photon via quantum frequency conversion}

\author{John Hannegan}
\author{James D. Siverns}
\affiliation{IREAP and the University of Maryland, College Park, Maryland 20742, USA}
\author{Qudsia Quraishi}
\affiliation{Army Research Laboratory, 2800 Powder Mill Road, Adelphi, Maryland 20783, USA}
\affiliation{IREAP and the University of Maryland, College Park, Maryland 20742, USA}
\homepage{https://quraishi.physics.umd.edu}
\date{\today}

\begin{abstract}
Future quantum networks will require the ability to produce matter-photon entanglement at photon frequencies not naturally emitted from the matter qubit.
This allows for a hybrid network architecture, where these photons can couple to other tools and quantum technologies useful for tasks such as multiplexing, routing, and storage, but which operate at wavelengths different from that of the matter qubit source, while also reducing network losses.
Here, we demonstrate entanglement between a trapped ion and a 780 nm photon, a wavelength which can interact with neutral-Rb-based quantum networking devices. A single barium ion is used to produce 493 nm photons, entangled with the ion, which are then frequency converted to 780 nm while preserving the entanglement. 
We generate ion-photon entanglement with fidelities $\geq$ 0.93(2) and  $\geq$ 0.84(2) for 493 nm and 780 nm photons respectively with the fidelity drop arising predominantly from a reduction in the signal-noise of our detectors at 780 nm compared with at 493 nm.
\end{abstract}

\maketitle

\section{\label{sec:Intro}Introduction}

Trapped ion systems are a leading platform in quantum computing \cite{zhu2021,blinov2021,Egan2021,HAFFNER2008}, simulation \cite{blatt2012,johanning2009,kyprianidis2021,Monroe2021} and short-distance (meter scale) quantum networking with both high rates and high entanglement fidelities \cite{Stephenson2020}. Trapped ion qubits are easily entangled with flying qubits for networking \cite{Blinov2004} and have demonstrated high fidelity single- and two-qubit gates \cite{Clark2021,Harty2014,Brown2011} combined with long trapping and coherence lifetimes \cite{Langer2005,Wu2021,wang2021}. As quantum networking nodes, barium ions, in particular, have advantageous properties that make them promising candidates with which to build these systems, including high qubit state preparation and measurement fidelities \cite{Christensen2020}, atomic transitions easily accessible with visible laser frequencies \cite{Hucul2017}, first-order magnetic field insensitive qubits in both ground states and metastable states \cite{Allcock2021} and the ability to implement multi-qubit gate operations with extremely low fundamental error rates over a wide range of laser frequencies \cite{Sawyer2021}.

Entanglement distribution in practical large-scale quantum networks will require both low-loss flying qubit transmission as well as the combination and integration of heterogeneous quantum systems, each with individual strengths, weaknesses, and unique uses.
Quantum frequency conversion (QFC) can be used to convert the frequency of a photon entangled with a quantum system to an optical frequency compatible with low loss fiber transmission \cite{Siverns2019,Krutyanskiy2019,bock2018high,Walker2018,Hannegan2022,Ikuta2016}.
This frequency freedom also provides the ability to integrate photonic devices \cite{Dong2022,saha2022}, and other, otherwise incompatible but useful classical and quantum systems \cite{Siverns2019slow,Craddock2019,Maring2017,Rakonjac2021} into an improved hybrid quantum network.

For the case of Ba$^+$, QFC allows for hybrid networking with neutral atom Rb systems \cite{Siverns2019slow,Craddock2019}.
Yet, to be useful for quantum networking, the frequency converted photon  must be shown to be entangled with the matter qubit.
Here, we use QFC to convert 493 nm photons entangled with a single $^{138}$Ba$^+$ ion to 780 nm, compatible with neutral Rb systems, in a way that preserves the polarization-based entanglement between the ion and the photon \cite{Ikuta2018}.
We characterize the ion-photon entanglement at both 493 nm and 780 nm, producing bounds on the observed entanglement fidelity and measure the rate of entangled photon production at both wavelengths.

%%%%%%%%%%%%%%%%%%%%%%%%%%%%%%%%%%%%%%%%%%%%%%%%%%%%5
\section{Experimental Details}
\begin{figure*}[htbp]
    %\centering
    \includegraphics[width=1\textwidth]{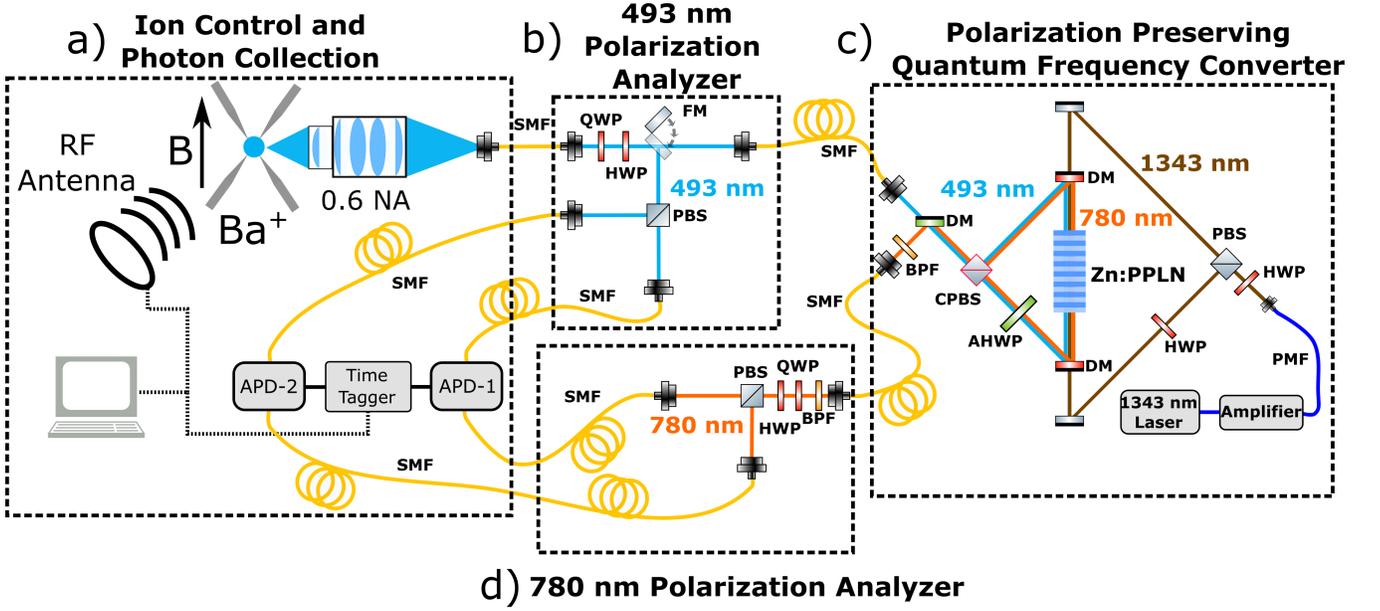}
    \caption{Experimental Layout. a) A $^{138}$Ba$^+$ ion produces 493-nm photons that are collected and coupled into a single mode fiber (SMF) by a 0.6 numerical aperture (NA) objective. The photon collection axis is orthogonal to the magnetic field at the ion. A radio-frequency (RF) antenna addresses the ion qubit based on photon measurements on avalanche photodiodes (APD-1 and APD-2).
    b) Photon polarization measurements at 493 nm are carried out via a polarization analyzer consisting of a quarter waveplate (QWP), half waveplate (HWP), and polarizing beamsplitter (PBS). A flipper mirror (FM) allows for the analyzer to be bypassed with the photons instead sent to the frequency conversion setup.
    c) Frequency conversion is performed using a waveguide written into Zn-doped periodically poled lithium niobate crystal (Zn:PPLN) at the center of two optical loops. Dichroic mirrors (DM) allow for combination and splitting of the multiple colors used. A custom-coated PBS (CPBS) and achromatic half waveplate (AHWP) operating at 493 nm and 780 nm enable these colors to share the same optical paths. A band-pass filter (BPF) is used to remove the majority of the noise photons produced by the 1343-nm pump light.
    d) A second polarization analyzer is used for 780-nm photonic qubit polarization measurements.}
    \label{fig:setup}
\end{figure*}

The experimental layout is outlined in Fig.~\ref{fig:setup}.
It consists of four concatenated setups, connected via a series of single-mode optical-fibers (SMF).
A $^{138}$Ba$^+$ ion produces 493-nm photons which are collected and coupled into a single-mode optical fiber using a 0.6 numerical-aperture (NA) lens system (Fig.~\ref{fig:setup}(a)). Photons coupled into the fiber are then either detected using a 493-nm polarization analyzer (Fig.~\ref{fig:setup}(b)) or sent to a quantum frequency converter (Fig.~\ref{fig:setup}(c)). 
If sent to the converter, the 493-nm photons are converted to 780-nm \cite{Siverns2019}, coupled into another SMF and detected via a 780-nm polarization analyzer (Fig.~\ref{fig:setup}(d)).
For entanglement measurements, operations on the ion are performed using a radio-frequency field emitted from an antenna placed close to the ion, and the measured ionic and photonic qubit states are used to determine bounds on the entanglement fidelity.

\subsection{Ion-Photon Entanglement Generation and Photon Collection}

\begin{figure}[htbp]
    \includegraphics[width=0.5\textwidth]{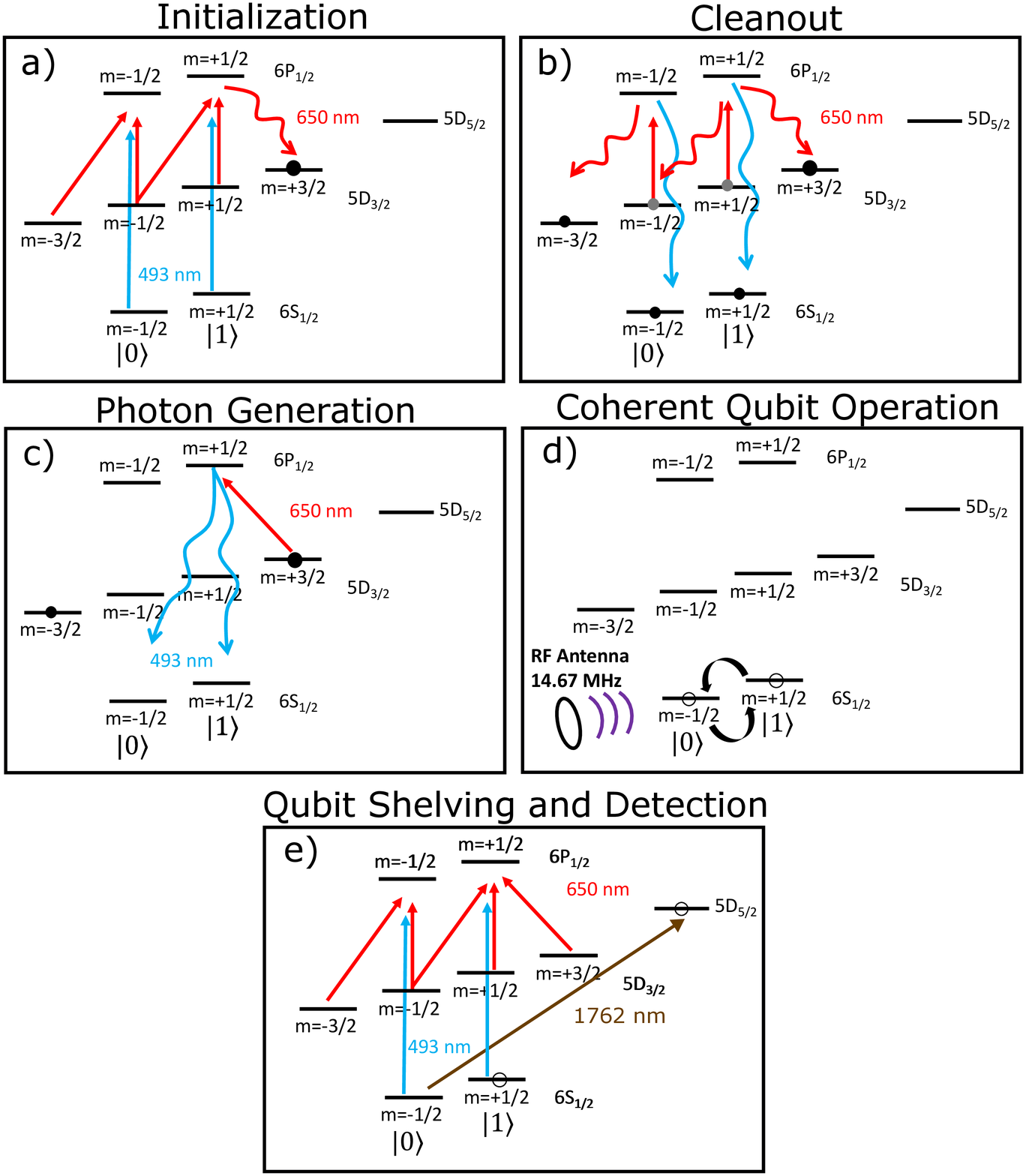}
    \caption{Photon production and ion qubit control. a) The ion is prepared into the $\ket{5D_{3/2},m_j=+1/2}$ state. 
    b) A clean out pulse is used to remove any remaining population in the $\ket{5D_{3/2},m_j=\pm1/2}$ states.
    c) The ion is excited to the $\ket{6P_{1/2},m_j=+1/2}$ state, from which a photon entangled with the ion is emitted.
    d) If a photon is detected, coherent ion-qubit operations are driven by an RF antenna tuned to 14.67 MHz.
    e) Ion-qubit state detection is performed via optical shelving of the $\ket{0}$ state via the $\Delta m = 0$ quadrupole transition at 1762 nm.}
    \label{fig:PhotonProduction}
\end{figure}

To generate photons entangled with qubit states in $^{138}$Ba$^+$, we follow the sequence outlined in Fig.~\ref{fig:PhotonProduction} and discussed in \cite{Siverns2017}.
We optically pump the ion into the $\ket{5D_{3/2},m_j=+3/2}$ edge state using an 8 $\mu$s exposure of $\pi$-polarized 493-nm light along with $\pi$ and $\sigma^+$-polarized 650-nm light (Fig.~\ref{fig:PhotonProduction}(a)).
A 1 $\mu$s exposure of $\pi$-polarized 650-nm light is then used to ensure any residual population left in $\ket{5D_{3/2},m_j=\pm1/2}$ is removed, thereby avoiding erroneous excitation in the next step (Fig.~\ref{fig:PhotonProduction}(b)).
A 200 ns pulse of $\sigma^-$-polarized 650-nm light then excites the ion to $\ket{6P_{1/2},m_j=+1/2}$, from where the ion can spontaneously decay to $\ket{6S_{1/2},m_j=\pm1/2}$, producing a single 493-nm photon with a temporal probability distribution related to the lifetime of the $\ket{6P_{1/2},m_j=+1/2}$ state.
A 10 $\mu$s period is then included after the 650-nm excitation pulse to allow our control system to time-tag (Fig.~\ref{fig:setup}~a)) detected photons.
We repeat this sequence in 500-attempt bursts, with each burst separated by 100 $\mu$s of Doppler cooling using all polarizations of 650-nm and 614-nm light and $\pi$-polarized 493-nm light. The 614 nm light is not shown in Fig. \ref{fig:PhotonProduction} for simplicity and is tuned to a frequency which allows pumping of population in the 5D$_{5/2}$ state to the 6P$_{3/2}$ state from where the ion can decay back into the  5D$_{3/2}$ or 6P$_{1/2}$ states.

Spontaneous emission of a 493-nm photon from the $\ket{6^2P_{1/2},m_j=+1/2}$ state in the above sequence produces a polarization-based ion-photon entangled state given by

\begin{equation}
    \ket{\Psi} = \sqrt{\frac{2}{3}}\ket{0}\ket{\sigma^+} + \sqrt{\frac{1}{3}}\ket{1}\ket{\pi},
\end{equation}

\noindent where we label the $\ket{6^2S_{1/2},m_j=-1/2}$ and $\ket{6^2S_{1/2},m_j=+1/2}$ ion states as $\ket{0}$ and $\ket{1}$, respectively. Taking into account the atomic radiation pattern, and assuming photon collection directly perpendicular to the quantization axis, defined by a 5.23 Gauss magnetic field produced by two permanent magnets, this entangled state reduces to

\begin{equation}
    \ket{\Psi} = \sqrt{\frac{1}{2}}\ket{0}\ket{V} + \sqrt{\frac{1}{2}}\ket{1}\ket{H},
    \label{eq:ionPhotonEnt}
\end{equation}

\noindent with $\sigma^+$-polarized and $\pi$-polarized photons projected to two orthogonal linear polarizations, denoted here by V and H \cite{Siverns2017,Crocker2019}.

We collect photons produced by the ion using a 0.6 NA objective (Special Optics: PN 15920), as shown in Fig.~\ref{fig:setup}~(a).
This objective focuses photons from the ion directly into an anti-reflection-coated single-mode fiber (Thorlabs: SM400).
In free space, the large solid angle of our photon collection would normally lead to significant and unavoidable polarization-based errors, as $\sigma^+$ and $\pi$-polarized photons are no longer orthogonal~\cite{Crocker2019}.
However, when coupling the collected light into a single-mode fiber, it has been shown that the non-orthogonal polarization components are cancelled out at the expense of collection efficiency \cite{Stephenson2020}, making Eq.~\ref{eq:ionPhotonEnt} accurate for our system.

The long temporal length of the 650-nm $\sigma^-$ excitation pulse (200 ns,) relative to the excited state lifetime of the P$_{1/2}$ manifold ($\approx 10$ ns), allows for multiple excitations of the ion in the event of decay back to the D$_{3/2}$ manifold ($\approx 25\%$ probability).
Decay to, and subsequent excitation from, the $\ket{5^2D_{3/2},m_j=+1/2}$ state to the $\ket{6^2P_{1/2},m_j=-1/2}$ state, can lead to spontaneous emission of a 493-nm photon that produces the ion-photon entangled state of Eq.~\ref{eq:ionPhotonEnt} with the photon polarization states swapped.
This state is orthogonal to the expected ion-photon entangled state, and will lead to a reduction in the measured entanglement fidelity \cite{Siverns2017,Crocker2019}.

To reduce the effect of errors due to these multiple excitations, we implement a software gate on the detected photon signal, referenced to the start of the 650-nm $\sigma^-$ excitation pulse.
We choose to only accept photon events occurring in a 40 ns window at the beginning of the photon's temporal profile, ignoring $\approx 17\%$ of the photon detection events which occur outside of this window. 
With the majority of photons produced via multiple excitations occurring towards the end of the photon's temporal profile (calculated via an optical Bloch equation model), this 40 ns software gate reduces the expected infidelity of the ion-photon entangled state due to multiple excitations from $\approx 9\%$ to $\approx 2\%$.

\subsection{Quantum Frequency Converter}\label{sec:QFC}

A simplified schematic of the setup used to convert the 493-nm photons emitted by the ion to 780-nm is shown in Fig.~\ref{fig:setup}~(c). 
A ridge waveguide in a Zinc-doped periodically poled lithium niobate crystal (Zn:PPLN) (NTT: WD-1344-000-A-C-C-TEC) facilitates difference frequency generation of S-polarized photons from 493-nm to 780-nm with the use of a high-intensity pump laser at 1343 nm \cite{Siverns2019,Siverns2019slow,Craddock2019}. As P-polarized photons are not efficiently converted \cite{klein2003absolute} by the Zn:PPLN, photons with this polarization must be rotated before being coupled into the waveguide.

To ensure the frequency conversion of both orthogonal polarizations of the photons emitted from the ion we place the Zn:PPLN waveguide at the center of two optical loops, each arranged in a Sagnac-interferometer-like configuration \cite{Ikuta2018}.
In the first loop (left-most loop of Fig.~\ref{fig:setup}~(c), single photons at 493 nm are collimated out of an SMF fiber and sent clockwise or anti-clockwise around the loop by transmission through, or reflection off of, a custom PBS (CPBS) operating at both 493 nm and 780 nm (Lambda Research Optics: BPB-10SF2-450-800).
In one direction of the loop, the 493-nm photons are rotated by an achromatic half waveplate (AHWP) (Thorlabs: AQWP05M-580) prior to entering the Zn:PPLN waveguide such that the photons' polarization is correct for difference frequency generation inside the Zn:PPLN waveguide. When travelling in the opposite direction of this loop the photons pass through the AHWP after passing through the Zn:PPLN waveguide.
In both loop directions these 493-nm photons are combined with 1343-nm light at dichroic mirrors (Thorlabs: DMLP950) and are coupled into the waveguide by  an off-axis parabolic mirror (Thorlabs: MPD00M9-P01, not shown).
Once converted to 780 nm, the photons are collimated by another parabolic mirror at the Zn:PPLN wavguide exit, recombined at the CPBS and then separated from pump and noise photons with a dichroic mirror for output fiber coupling.
A 10 nm wide bandpass filter (Thorlabs: FBH780-10), is included just before the output fiber to filter out any pump light present as well as the majority of noise photons produced through other nonlinear processes involving the pump laser \cite{Pelc11}.

In the second (right-most of Fig.~\ref{fig:setup}~(c)) loop, amplified (MPB: RFA-P-5-1341-SF) 1343-nm pump light is split, based on polarization at a PBS, and coupled into the waveguide in the same manner as described above.
A half waveplate (HWP) before this PBS, combined with control of the output power of the 1343-nm amplifier, allows for arbitrary control over the power sent in each direction of the loop.
This allows the conversion efficiency of each direction of the loop to be independently tuned and matched.

\begin{figure}[htbp]
    \includegraphics[width=0.5\textwidth]{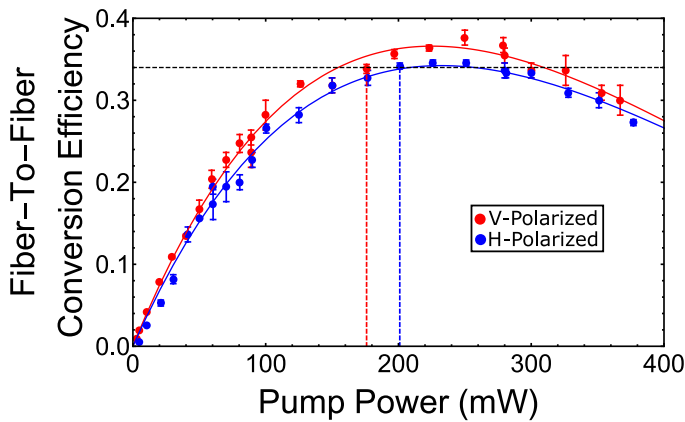}
    \caption{Single photon quantum frequency conversion efficiencies for each polarization, measured from the output of the 493-nm SMF fiber in Fig.~\ref{fig:setup}~(c) to the output of the 780-nm fiber in Fig.~\ref{fig:setup}~(d).
    The horizontal dashed line represents the conversion efficiency used in this experiment ($34.5\%$).
    The vertical red and blue lines represent the pump power needed to reach this conversion efficiency for vertical and horizontal input photon polarizations, respectively.
    }
    \label{fig:ConversionEfficiency}
\end{figure}

The input-fiber-to-output-fiber conversion efficiency of the quantum frequency converter as a function of pump power is shown in Fig.~\ref{fig:ConversionEfficiency}, for each polarization of input 493-nm light.
We measure peak conversion efficiencies of 37.9(9)$\%$ and 34.5(4)$\%$ for V and H polarizations, respectively.
The pump powers in each direction are set to match the conversion efficiency for each polarization to the minimum of these two values, at 34.5$\%$. 
These power settings are represented by the vertical dashed lines in Fig.~\ref{fig:ConversionEfficiency}.
At the operational conversion efficiency shown in Fig.~\ref{fig:ConversionEfficiency}, we measure $\approx$ 200 counts/s noise on each detector above their dark count level.

Using the described conversion scheme, the converted 780-nm photon should retain the same entanglement with the ion as the 493-nm photon, given by Eq.~\ref{eq:ionPhotonEnt}, except with the polarization state of the photon swapped from H to V and vice-versa.
This polarization swapping is due to the AHWP (Fig. \ref{fig:setup}(c)) placed in the 493/780-nm loop through which each polarization travels once. If so desired a second HWP placed just before the 780-nm fiber could be used to rotate the polarization back to the original state.

\subsection{Qubit Measurement and Manipulation}

Determination of the ion-photon entanglement fidelity requires the ability to perform coherent operations on both the trapped ion and photonic qubits, as well as make measurements on both qubits in different bases.
We perform basis rotations and measurements on the photonic qubit at the polarization analyzers shown in Fig.~\ref{fig:setup}, depending on the color of the photon.
We perform operations on the ion by directly driving coherent qubit transitions using radio-frequency (RF) pulses tuned to qubit resonance.
Operations on the ion must be carefully timed relative to photon emission in order to ensure the proper phase of these gates. 

\subsubsection{Polarization Analyzers}

The polarization analyzers shown in Figs.~\ref{fig:setup}(b) and (d) consist of both a quarter waveplate (QWP) and half waveplate (HWP), each mounted in a motorized rotation mount (Thorlabs: DDR25) with rotation angle controlled by our FPGA control system \cite{Artiq}.
We use a flipper mirror (FM) on the 493 nm polarization analyzer (Fig.~\ref{fig:setup}(b)) to selectively route 493-nm photons to measurement or to the frequency converter and 780-nm analyzer.
The waveplates allow us to transform any arbitrary photon polarization to a linear polarization, enabling us to undo any polarization rotations caused by birefringence present in the fiber(s) before each analyzer.
Polarization-based photon measurements are then made by the polarizing beam splitter cube and single photon detecting avalanche photodiodes (APD) (PerkinElmer: SPCM-AQR-15-FC), allowing simultaneous measurement of both orthogonal photon polarizations.
This configuration allows for measurements in the horizontal-vertical ($\{H,V\}$) and diagonal-anti-diagonal ($\{D,A\}$) photonic qubit bases, sufficient for determination of upper and lower bounds on the ion-photon entanglement fidelity \cite{Blinov2004,auchter2014}.

\subsubsection{Ion Control and Timing}

The radio-frequency (RF) pulses used for ion qubit rotations are provided by the RF antenna shown in Fig.~\ref{fig:setup}(a). The RF signal, produced by a direct digital synthesizer (DDS) (AD9910) connected to the experimental control system \cite{Artiq}, is set on resonance with the ion qubit splitting at 14.67 MHz.
We perform RF rotations at a fixed time delay relative to the recorded single photon detection time tags.
Additionally, we use these time tags to actively program the DDS such that the phase of the applied RF is synchronous with the free evolution of the ion qubit, with a controllable offset dependent on the desired gate to be performed.
Due to the timing delay required to program the phase and frequency of the DDS after photon detection ($\approx 70 \mu$s), an additional spin-echo pulse is applied to the ion as a part of these operations to reduce ion-qubit dephasing effects due to local magnetic field noise \cite{ruster2016long,szwer2010keeping}. This spin-echo extends our qubit coherence time from $\approx$ 200 $\mu$s to $\approx$ 2 ms.

We perform state detection on the ion by optically shelving the $\ket{0}$ state to the $\ket{D_{5/2},m_j=-1/2}$ state using a quadrupole transition at 1762 nm (Fig. \ref{fig:PhotonProduction}(e)) \cite{dietrich2010hyperfine,Hannegan2022thesis}.
This is then followed by 3 ms of fluorescence detection on the ion using light at 493 nm and 650 nm.
After state detection, we illuminate the ion with laser light at 614 nm to remove it from the $\ket{D_{5/2},m_j=-1/2}$ state and allow the photon production sequence to recommence.

%%%%%%%%%%%%%%%%%%%%%%%%%%%%%%%%%%%%%%%%%%%%%5
\section{Results}

\begin{figure*}[htbp]
    \includegraphics[width=1\textwidth]{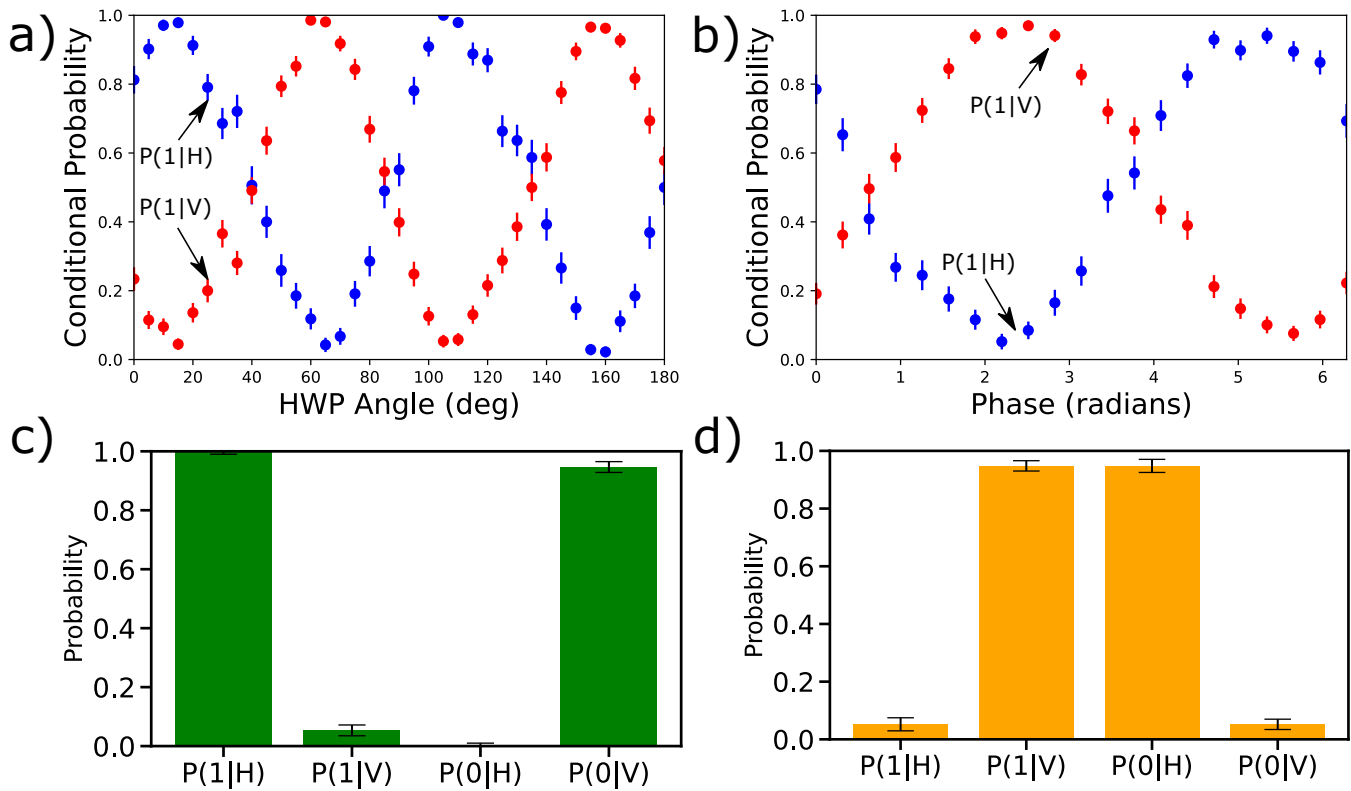}
    \caption{Ion-photon correlations at 493 nm.  a) Calibration scans for the unrotated basis (z-basis) where the blue(red) data represents the probability of detecting the ion in state $\ket{1}$ given that the photon is detected in state $\ket{H}(\ket{V})$ by APD-1(APD-2) for a given position of the HWP in Fig.~\ref{fig:setup}~b. b) Calibration scans for rotated basis (x-basis), obtained by setting the HWP in Fig.~\ref{fig:setup}~(b) to $40^{\circ}$, and for a given phase of the $\pi/2$ pulse applied to the ion.
    c) Conditional measurement probabilities at the point of maximum correlations ($105^{\circ}$) for the unrotated basis data in a).
    d) Conditional measurement probabilities at the point of maximum correlations (7$\pi/10$ radians) for the rotated basis data in b).
    All error bars are statistical, with 500 photon events for each data point.}
    \label{fig:493Correlations}
\end{figure*}

We can calculate a lower bound on the fidelity for our ion-photon entangled state at 493-nm by calculating \cite{Blinov2004}

\begin{equation}
\begin{split}
    F_{493} \geq \frac{1}{2}(\rho_{H1,H1} + \rho_{V0,V0} - 2\sqrt{\rho_{H0,H0}\rho_{V1,V1}}\\+ \tilde{\rho}_{H1,H1} +  \tilde{\rho}_{V0,V0} -  \tilde{\rho}_{H0,H0} -  \tilde{\rho}_{V1,V1}).
\end{split}
\label{eq:493LowerBound}
\end{equation}

The matrix elements $\rho_{\gamma b,\gamma b}$ (where $\gamma$ is the photon polarization and $b$ is the ion state) are given by P$(\gamma)$P$(b\vert \gamma)$, where  P$(\gamma)$ is the total probability of detecting a photon with polarization $\gamma$ and P$(b\vert\gamma)$ represents the conditional probability of measuring the ion in state $b$ given a measurement of the photon's polarization as $\gamma$.
The matrix elements $\tilde{\rho}_{\gamma b,\gamma b}$ are given with the same form but after a rotation by a polar angle of $\pi/2$ on the Bloch sphere of both the photon and ion qubits.
In a similar manner, we can also calculate an upper bound on the fidelity as \cite{auchter2014}

\begin{equation}
\begin{split}
    F_{493} \leq \frac{1}{2}\left(\sqrt{\rho_{H1,H1}} + \sqrt{\rho_{V0,V0}}\right)^2.
\end{split}
\label{eq:493UpperBound}
\end{equation}

As described in Section \ref{sec:QFC}, we measure the 780-nm ion-photon entanglement with the horizontal and vertical polarizations swapped compared with the 493-nm ion-photon entanglement, such that a lower bound is given by

\begin{equation}
\begin{split}
    F_{780} \geq \frac{1}{2}(\rho_{H0,H0} + \rho_{V1,V1} - 2\sqrt{\rho_{H1,H1}\rho_{V0,V0}}\\ + \tilde{\rho}_{H0,H0} +  \tilde{\rho}_{V1,V1} -  \tilde{\rho}_{H1,H1} -  \tilde{\rho}_{V0,V0}),
\end{split}
\label{eq:780LowerBound}
\end{equation}

\noindent and the upper bound is given by

\begin{equation}
\begin{split}
    F_{780} \leq \frac{1}{2}\left(\sqrt{\rho_{H0,H0}} + \sqrt{\rho_{V1,V1}}\right)^2.
\end{split}
\label{eq:780UpperBound}
\end{equation}

This choice is somewhat arbitrary in theory, as the half waveplate in either polarization analyzer (Fig. \ref{fig:setup}(b) and (d)) can be used to swap $H \leftrightarrow V$.

\subsection{Entanglement Fidelity Measurements}
We first perform calibrations of the 493-nm photonic qubit measurement basis.
This is performed by measuring the conditional probabilities, P$(b\vert\gamma)$, for varying rotation-angles of both the quarter and half waveplates in the 493-nm polarization analyzer (Fig. \ref{fig:setup}(b)).
We optimize the quarter waveplate angle to provide the maximum ion-photon correlation visibility, $\vert $P$(1\vert H)-$P$(1\vert V)\vert$, when scanning the half waveplate angle.
The 493-nm ion-photon correlations for this optimized quarter waveplate angle are shown in Fig.~\ref{fig:493Correlations}~(a) as a function of the half-waveplate angle. There are 500 detection events measured at each half-waveplate angle.

We then measure correlations between the ion and photonic qubit states following rotation of both qubits by $\pi/2$ on the Bloch sphere.
With the quarter waveplate position optimized as described above, we apply a $\pi/2$ rotation to the photon polarization by setting the half waveplate to an angle of $40^{\circ}$ (Fig.~\ref{fig:493Correlations}~a)).
We rotate the ion measurement basis using a $\pi/2$ RF pulse with a set phase and a fixed delay time relative to the photon detection time tag. 
By scanning this phase, we scan the relative phase of rotation of the ion versus that of the qubit, again measuring correlations between the ion qubit state and the polarization of the photon.
This results in the rotated-basis correlation fringes shown in Fig.~\ref{fig:493Correlations}~(b), with 500 events measured at each phase of the applied RF.

\begin{figure*}[htbp]
    \includegraphics[width=1\textwidth]{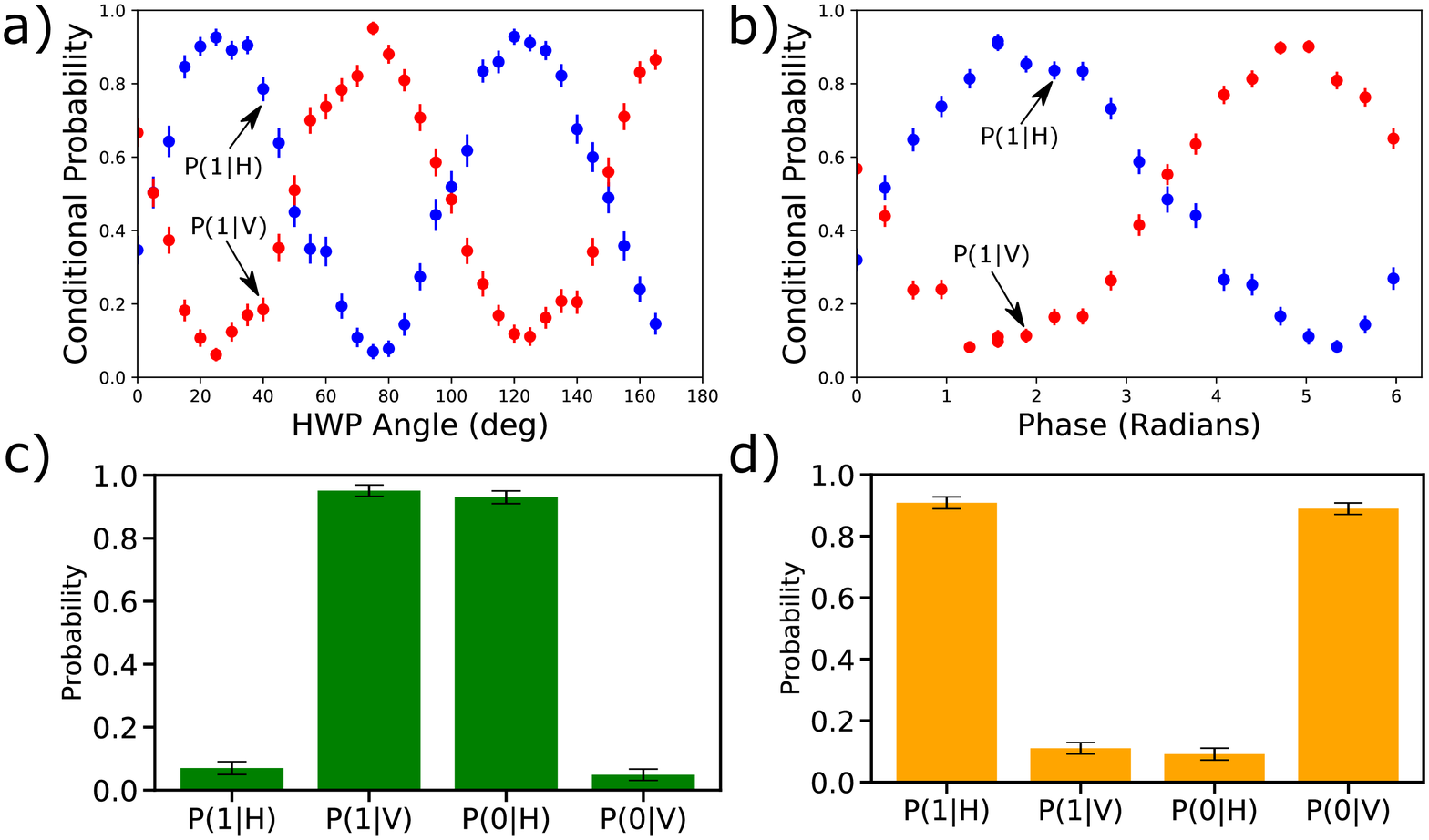}
    \caption{Ion-photon correlations at 780 nm.  a) Calibration scans for the unrotated basis (z-basis) where the blue(red) data represents the probability of detecting the ion in state $\ket{1}$ given that the photon is detected in state $\ket{H}(\ket{V})$ by APD-1(APD-2) for a given position of the HWP in Fig.~\ref{fig:setup}~(b). b) Calibration scans for rotated basis (x-basis), obtained by setting the HWP in Fig.~\ref{fig:setup}~(b) to $100^{\circ}$, and for a given phase of the $\pi/2$ pulse applied to the ion.
    c) Conditional measurement probabilities at the point of maximum correlation ($75^{\circ}$) for the unrotated basis data in (a).
    d) Conditional measurement probabilities at the point of maximum correlation ($\pi/2$ radians) for the rotated basis data in (b).
    All error bars are statistical, with 500 photon events for each data point.}
    \label{fig:780Correlations}
\end{figure*}

At the point of maximum visibility in the unrotated basis, ($105^\circ$ in Fig.~\ref{fig:493Correlations}(a)), we measure P$(H\vert1)=1.00(1)$, P$(V\vert0)=0.95(2)$, P$(H\vert0)=0.00(1)$, and P$(V\vert1)=0.05(2)$, as shown in Fig.~\ref{fig:493Correlations}~(c).
Similarly, in the rotated basis (Fig.~\ref{fig:493Correlations}~(b)), we measure $\tilde{P}(H\vert0)=0.95(2)$, $\tilde{P}(V\vert1)=0.94(2)$, $\tilde{P}(H\vert1)=0.05(2)$, and $\tilde{P}(V\vert0)=0.06(2)$, at the point of maximum correlation (7$\pi/10$ radians).
From these results, we use Eq.~\ref{eq:493LowerBound} and Eq.~\ref{eq:493UpperBound} to calculate bounds on the ion-photon entanglement at 493 nm, $0.93(2)\leq  F_{493}\leq 0.96(2)$.

After measurement of ion-photon entanglement at 493 nm, we set the flipper mirror in Fig. \ref{fig:setup}(b) to send the single photons into the quantum frequency conversion setup and 780 nm polarization analyzer.
We use the same waveplate optimization procedure as described for the 493-nm data to optimize the 780-nm quarter waveplate position (Fig. \ref{fig:setup}(d)), resulting in the optimized 780-nm ion-photon correlations shown in Fig.~\ref{fig:780Correlations}~(a).
By setting the half waveplate to $100^\circ$ and using the same $\pi/2$ RF pulse procedure on the ion as described above, we measure correlations in the rotated basis, the results of which are shown in Fig.~\ref{fig:780Correlations}~(b). The results in each basis are taken using 500 photon detection events for each data point.

For the 780-nm data, we measure, at the point of maximum correlation ($75^{\circ}$ in Fig.~\ref{fig:780Correlations}~(a)), P$(H\vert0)=0.93(2)$,  P$(V\vert1)=0.95(2)$, P$(H\vert1)=0.07(2)$, and P$(V\vert0)=0.05(2)$ for the unrotated basis, shown in Fig.~\ref{fig:780Correlations}~(c).
At the point of maximum visibility in the rotated basis ($\pi/2$ radians in Fig.~\ref{fig:780Correlations}~(b)), we measure $\tilde{P}(H\vert1)=0.91(2)$, $\tilde{P}(V\vert0)=0.89(2)$, $\tilde{P}(H\vert0)=0.09(2)$, and $\tilde{P}(V\vert1)=0.11(2)$, as represented by Fig.~\ref{fig:780Correlations}~(d).
Using these values in Eq.~\ref{eq:780LowerBound} and Eq.~\ref{eq:780UpperBound}, we find $ 0.84(2)\leq F_{780}\leq 0.94(2)$.

\subsection{Sources of Infidelity}
The measured lower bound of the entanglement fidelity for both photon wavelengths is well above the classical limit of $F>0.5$. 
The fidelity is reduced from unity at both wavelengths by several experimental factors, summarized in Table~\ref{tbl:Errors}.
Common to both measurements is infidelity caused by imperfect state detection ($1.5\%$), photon production ($1.5-2\%$), ion qubit rotation using the RF signal ($1\%$). 
The reduced fidelity at 780 nm, compared to that at 493 nm, can be attributed to the two following factors. 
First, a reduced signal-to-noise ratio (SNR) at 780-nm (SNR $\approx$ 10) compared to that at 493-nm (SNR $\approx$ 55) leading to infidelities in the entangled state of $6\%$ and $1.2\%$ respectively. 
Second, an increase in polarization rotation and measurement errors at 780 nm (causing $\approx1-5\%$ infidelity) compared to the 493 nm measurement (causing $\approx1-3\%$ infidelity). 
The infidelity caused by polarization rotation errors can be attributed to drifts in fiber birefringence over the course of the experiment(s) and from waveplate rotation errors causing non-optimal calibration of the photon polarization basis measurement.
Improvements in detector efficiency ($\approx 58\%$ at 780 is used in this work with $>90\%$ commercially available), along with narrower noise filtering of the DFG output can help to reduce errors by acting to increase the SNR. The improved signal rates combined with reduced noise rates will result is shorter experimental run times leading to less drift in fiber birefringence over the experiment run-time and, therfore, higher measured ion-photon fidelity.

\begin{table*}[!htb]
\begin{center}
    \begin{tabular}{ | c | c| c |}
        \hline
        &Entanglement Infidelity& Entanglement Infidelity\\
        Error Source & 493 nm  & 780 nm \\
        \hline\hline
         State Detection & 1.5 & 1.5\\
        \hline
        Photon Production & 1.5-2.0 & 1.5-2.0\\
        \hline
        Polarization Rotation and Measurement & 1-3 & 1-5\\
        \hline
        Signal-to-Noise Ratio & 1.2 & 6 \\
        \hline
        RF Gate Errors and Qubit Decoherence & 1 & 1\\
        \hline
        Sum of Infidelities & 6.2-8.7 & 11 - 15.5\\
        \hline
    \end{tabular}
\end{center}
    \caption{Summary of experimental imperfections and their contribution to the reduction in the measured fidelity at both 493 nm and 780 nm. Values are given in units of percent. The range of values given for certain values are a result of experimental uncertainties or are due to observed drifts throughout the experimental runtime.}
    \label{tbl:Errors}
\end{table*}
\subsection{Entanglement Rates}

\begin{figure}[htbp]
    \includegraphics[width=0.5\textwidth]{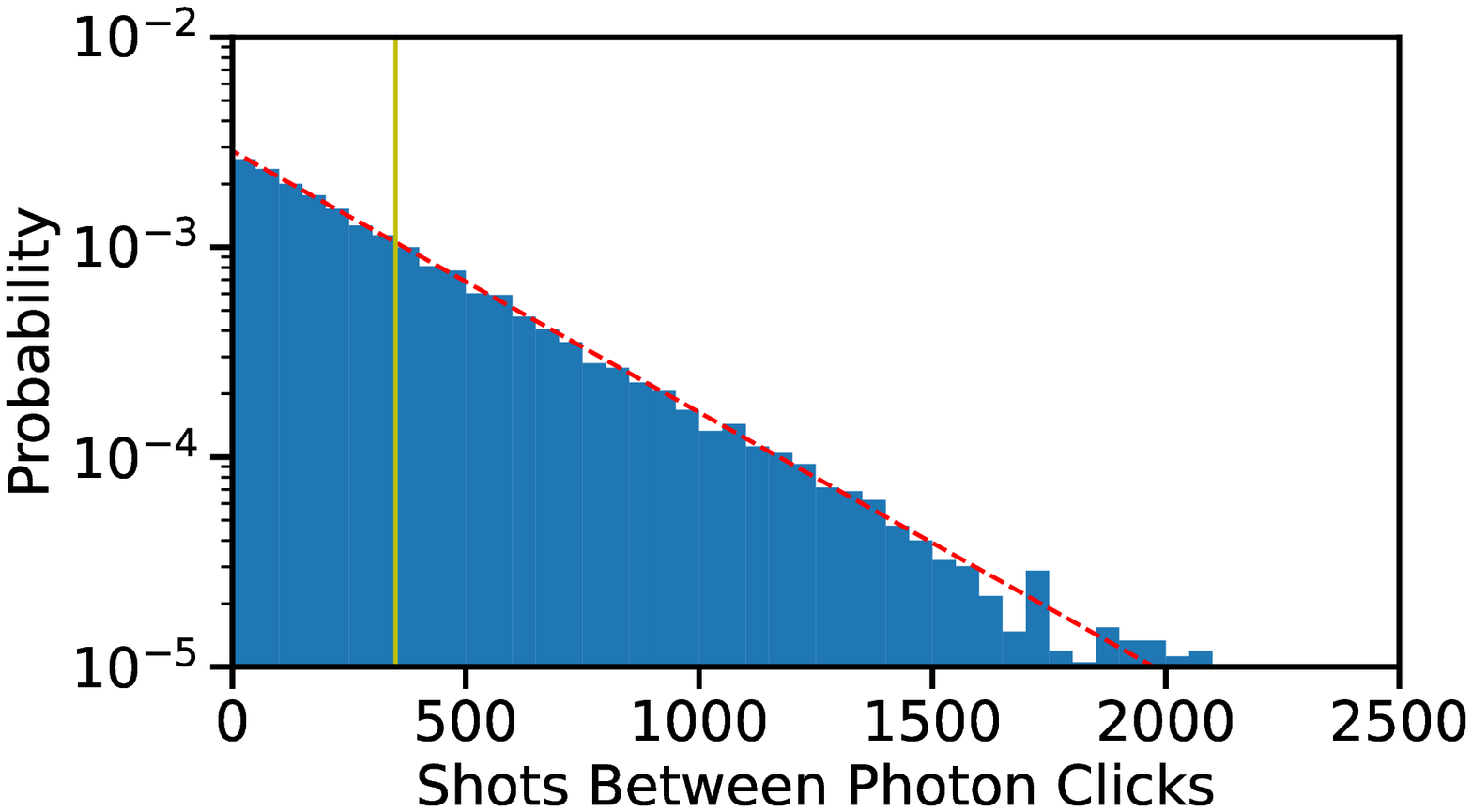}
    \caption{Histograms of photon production attempts between successful detection at 493 nm, using 28500 total detection events.
    The exponential fit (red dashed line) suggests a mean (vertical line) of 350 attempts before a successful photon detection.
    This corresponds to a entangled photon production rate of 137 s$^{-1}$ during the fast loop of the experiment.}
    \label{fig:493Rate}
\end{figure}

\begin{figure}[htbp]
    \includegraphics[width=0.5\textwidth]{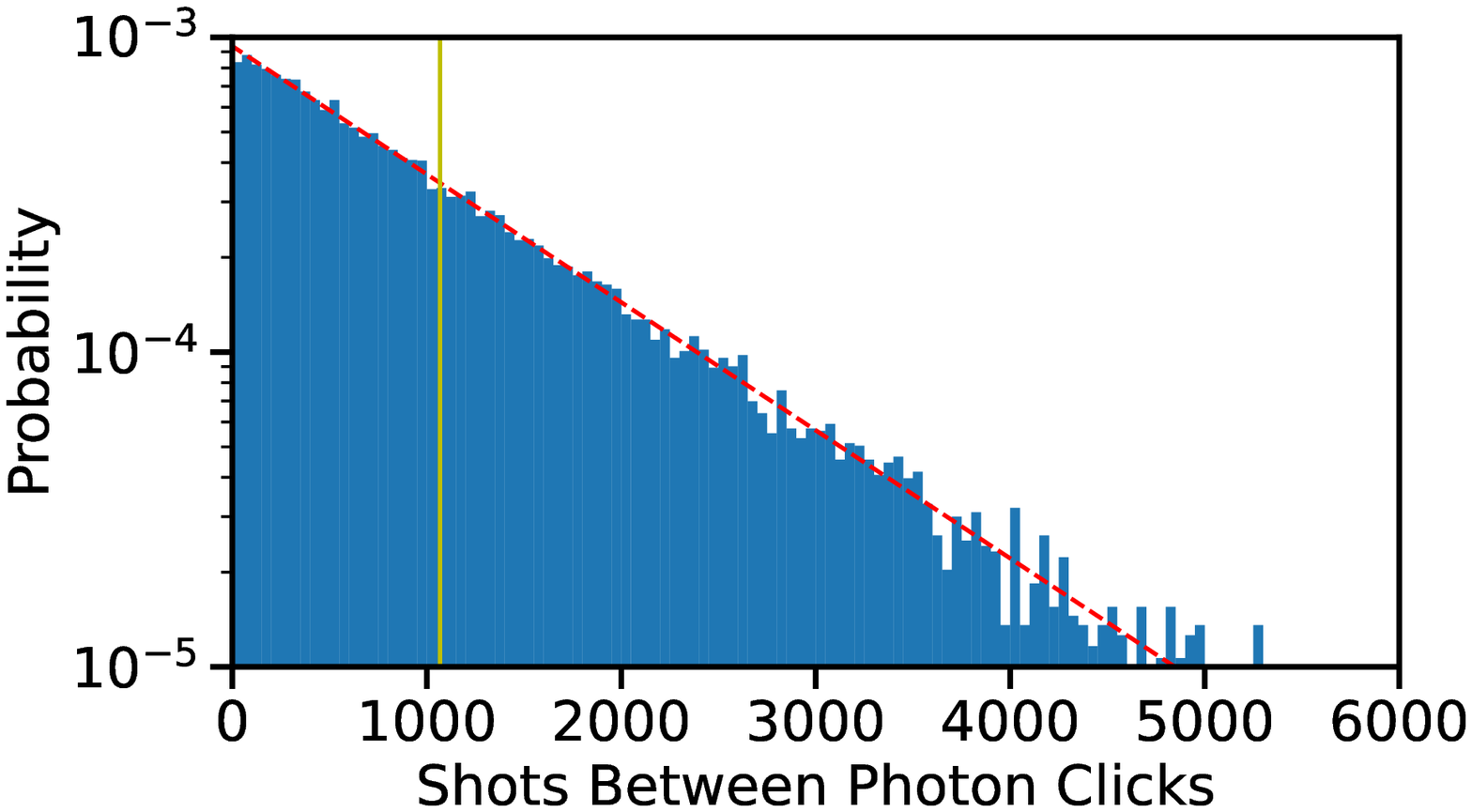}
    \caption{Histograms of photon production attempts between successful detection at 780 nm, using 20700 total detection events.
    The exponential fit (red dashed line) suggests a mean (vertical line) of 1068 attempts before a successful photon detection.
    This corresponds to a entangled photon production rate of 45 s$^{-1}$ during the fast loop of the experiment.}
    \label{fig:780Rate}
\end{figure}

In addition to the fidelity of the entanglement, we also make measurements of the generation rate of the observed ion-photon entanglement.
By recording the number of experimental cycles between successive photon detection events during data collection, we produce the histograms shown in Fig.~\ref{fig:493Rate} and Fig.~\ref{fig:780Rate}, for the 493 nm and 780 nm data, respectively.
Exponential distribution fits to these data (dashed red lines) give an average of 350 photon production attempts between successful detection events at 493 nm (vertical line Fig.~\ref{fig:493Rate}), with an average of 1068 attempts between successful detection events at 780 nm (vertical line Fig.~\ref{fig:780Rate}).
With an photon attempt repetition rate of $\approx 48\times10^3$ attempts/s this results in an average ion-photon entanglement generation rate of 143 events/s at 493 nm and 47 events/s at 780 nm.
This entanglement rate at 780 nm is comparable with other trapped ion systems using a single stage of quantum frequency conversion \cite{bock2018high}.

These rates can be improved by reducing optical losses between each of the concatenated setups ($\approx 66\%$ transmission), increased photon detection efficiency ($>90\%$ is possible at both 493 nm and 780 nm compared to the current values of $\approx 40\%$ and $\approx 58\%$), and better optimized fiber coupling using the 0.6 NA lens (currently $\approx$ 38$\%$). Additionally, an increase in photon attempt rate would give the largest gains to entanglement rate. 
We are currently limited by a required 10 $\mu$s delay every loop to allow the control system to read-in photon time-tags and perform conditional logic to decide if a photon has been detected or not \cite{Artiq}. 
This however, has been shown to be performed in well under a microsecond with customized hardware \cite{Stephenson2020}, which would immediately increase our photon production attempt rate by a factor of $\approx 2$. 
A further increase in repetition rate could be achieved through use of a pulsed 493-nm laser, enabling faster state preparation and photon extraction in a manner similar to \cite{Stephenson2020}, which could provide repetition rates approaching 1 MHz, or a factor of $\approx$ 20 improvement in entangled photon rate.

\section{Conclusion}
In summary, we have demonstrated the entanglement of 780-nm photons with a trapped $^{138}$Ba$^+$ ion, obtained through polarization-preserving quantum frequency conversion of the ion's native 493 nm emission.
The ion-photon entanglement fidelity at 780 nm is bounded at $F\geq0.84(2)$, well above the classical limit of $F\geq0.5$, and may further be improved through increased filtering of noise produced in the quantum frequency process combined with a reduction in fiber-based polarization drifts.
The demonstrated entanglement rate at 780 nm is comparable to similar trapped ion systems using quantum frequency conversion \cite{bock2018high}, but may be improved by up to a factor of $\approx 20$ through improvements to our control system and experimental apparatus. 
Nevertheless, the demonstrated rates and fidelites can enable experimental investigations into hybrid quantum networks consisting of both trapped Ba$^+$ ions and neutral Rb atoms as well as extending the networking range of Ba$^+$ ions from a few meters to a few kilometers. 
This could enable direct entanglement between these two fundamentally different platforms \cite{Craddock2019}, as well as allow investigations into improving trapped-ion-based quantum network using photon detection and storage techniques based on neutral atom systems \cite{Hannegan2021}.

\begin{acknowledgments}
All authors acknowledge support from the United States Army Research Lab's Cooperative Agreement at the University of Maryland and the Army Research Lab.
\end{acknowledgments}

\bibliography{ent}

\end{document}